\documentclass[9pt,twocolumn,twoside]{osajnl}

\journal{ol} 

\setboolean{shortarticle}{true}

\title{Topological edge states of nonequilibrium polaritons in hollow honeycomb arrays}

\author[1,*]{Xuekai Ma}
\author[2,3,4]{Yaroslav V. Kartashov}
\author[5]{Albert Ferrando}
\author[1,6]{Stefan Schumacher}

\affil[1]{Department of Physics and Center for Optoelectronics and Photonics Paderborn (CeOPP), Universit\"{a}t Paderborn, Warburger Strasse 100, 33098 Paderborn, Germany}
\affil[2]{ICFO-Institut de Ci{\`e}ncies Fot{\`o}niques, The Barcelona Institute of Science and Technology, 08860 Castelldefels (Barcelona), Spain}
\affil[3]{Institute of Spectroscopy, Russian Academy of Sciences, Troitsk, Moscow, 108840, Russia}
\affil[4]{Russian Quantum Center, Skolkovo 143025, Russia}
\affil[5]{Interdisciplinary Modeling Group, Departament d’{\`O}ptica, Universitat de Val{\`e}ncia, Doctor Moliner 50, E-46100 Burjassot (Val{\`e}ncia), Spain}
\affil[6]{College of Optical Sciences, University of Arizona, Tucson, AZ 85721, USA}

\affil[*]{Corresponding author: xuekai.ma@gmail.com}

\begin{abstract}
We address topological currents in polariton condensates excited by uniform resonant pumps in finite honeycomb arrays of microcavity pillars with a hole in the center. Such currents arise under combined action of the spin-orbit coupling and the Zeeman splitting that break the time-reversal symmetry and open a topological gap in the spectrum of the structure. The most representative feature of this structure is the presence of two interfaces, inner and outer ones, where the directions of topological currents are opposite. Due to the finite size of the structure polariton-polariton interactions lead to the coupling of the edge states at the inner and outer interfaces, which depends on the size of the hollow region. Moreover, switching between currents can be realized by tuning the pump frequency. We illustrate that currents in this finite structure can be stable and study bistability effects arising due to the resonant character of the pump.
\end{abstract}
\setboolean{displaycopyright}{true}

\begin{document}

\maketitle
The attention to new topological phases of matter has grown dramatically during the last decade, fueled mainly by potential applications of topologically protected edge states, which exist in such materials and demonstrate unusual propagation properties, resistant to disorder and inhomogeneities, and the ability to traverse even sharp material bends and corners. Discovered first in electronic systems \cite{elect1,elect2}, topological insulators were predicted and successfully demonstrated in diverse areas of science, including photonic and optoelectronic systems \cite{photon1,photon2}. Multiple approaches to realization of photonic topological phases are known, some of which make use of gyromagnetic photonic crystals \cite{giro1,giro2}, coupled microresonator arrays \cite{reson1,reson2}, modulated Floquet \cite{floq1,floq2} and many other structures. While in hybrid polaritonic systems technologically fabricated micropillar arrays \cite{toppol1,toppol2} or lattices induced by acoustic waves \cite{toppol3} can support topological currents, the predictions recently culminated in observation of polariton topological insulators in two- \cite{toppol4} and one-dimensional \cite{toppol5} geometries. 

\begin{figure}[!b]
\centering
{\includegraphics[width=1\linewidth]{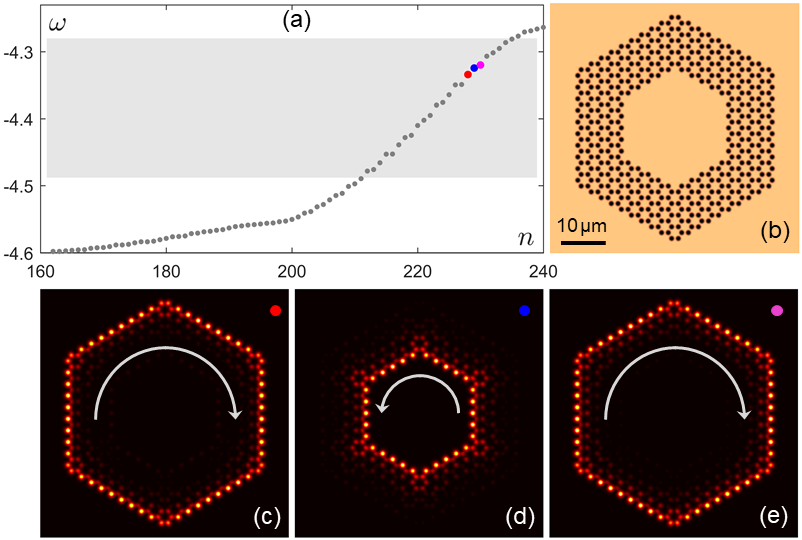}}
\caption{\textbf{Eigenstates of a hollow honeycomb array.} (a) Eigenfrequencies $\omega_n$ of linear modes versus mode index $n$. The gray area represents the band gap of the bulk array, where only the edge states appear. (b) Hollow honeycomb array. (c-e) $|\Psi_{-}|$ distributions corresponding to points of different color in (a). White arrows indicate the propagation direction of polaritons.}
\label{fig:EigenStates}
\end{figure}

Polariton topological insulators are intrinsically nonlinear systems, where polariton-polariton interactions play an important role in the condensation process. This perfectly fits into modern trends in topological photonics that now turns toward investigation of nonlinear effects, see a recent review \cite{nonltop1}. Nonlinearity brings a number of new effects in topological systems, among which are the formation of topologically protected solitons \cite{topsol1,topsol2,topsol3,topsol4,topsol5}, predicted for polaritons in \cite{topsol6,topsol7,topsol8}, nonlinearity-induced inversion of topological currents \cite{nonlpol1}, topological phases induced by vortex lattices \cite{nonlpol2}, coupling between corner modes in higher-order polariton insulators \cite{nonlpol3}, as well as bistability \cite{bistab1,bistab2}, all of which considerably extend the tools for the control of topologically protected currents. One of the most convenient platforms for the investigation of nonlinear effects in polariton condensates is offered by the arrays of microcavity pillars \cite{arrays1,arrays2}, which can be used for the creation of insulators of various symmetries, from conventional honeycomb \cite{toppol4,honey1}, to Lieb \cite{topsol8,lieb1}, and kagome \cite{topsol7,kagome1} 
ones.

In this Letter we investigate the impact of nonlinearity on the properties of topological edge currents in polariton condensates in hollow honeycomb arrays, which simultaneously feature two interfaces - inner and outer ones. We show that a resonant uniform pump allows selective excitation of the topological currents either at one or at both interfaces, and that nonlinearity may couple two states at the opposite interfaces due to their proximity, thereby offering the control over shape and determining stability of the currents.

\begin{figure}[t]
\centering
{\includegraphics[width=1\linewidth]{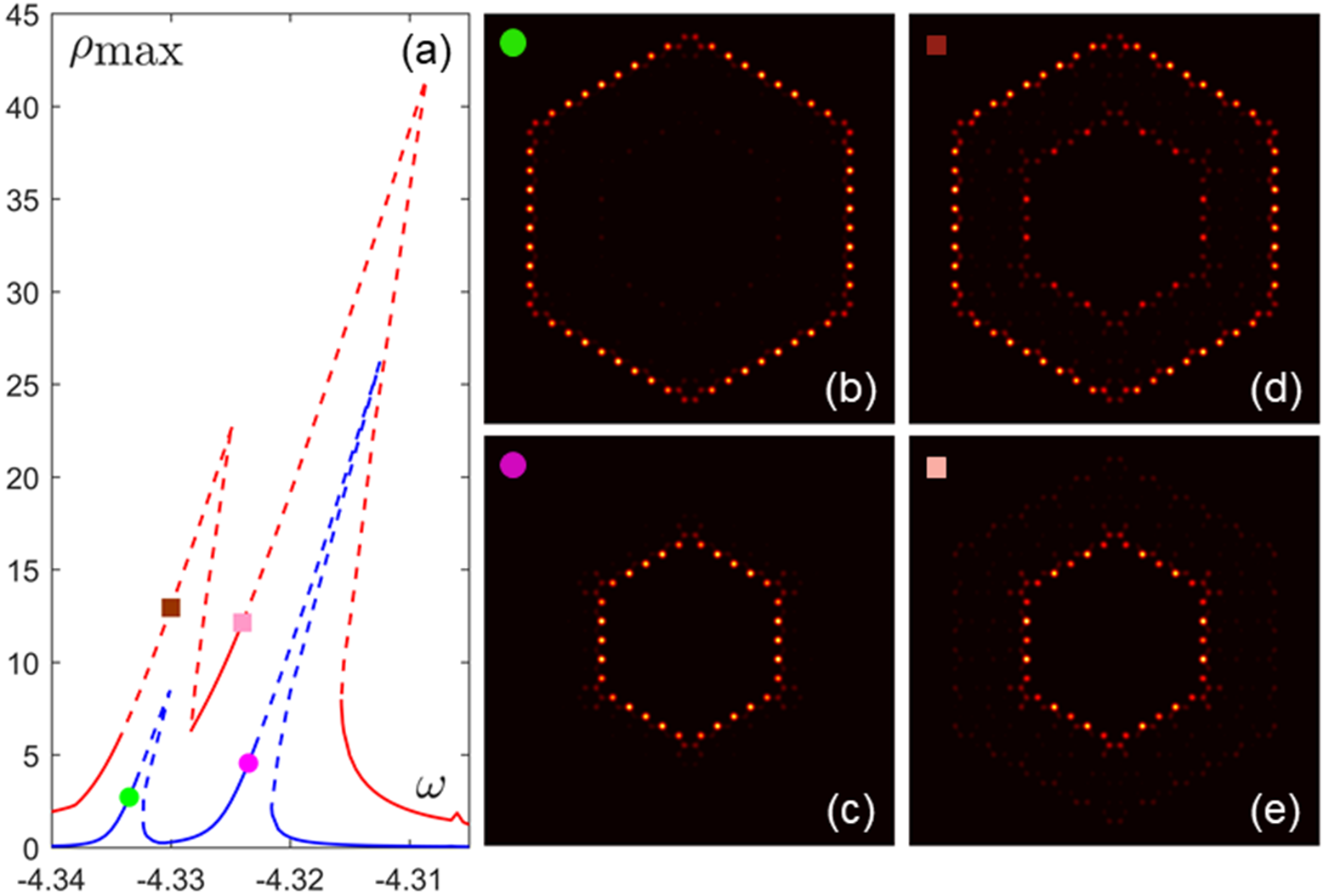}}
\caption{\textbf{Edge states in the nonlinear regime.} (a) Dependence of the maximum density ($\rho_\text{max}$ in $\mu$m$^{-2}$) of $\Psi_{-}$ on the frequency of the pump ($\omega$ in THz) for different pump amplitudes: $E_{\pm}^{0}=$ 0.00002 (blue lines) and 0.0001 (red lines). Solid lines are stable solutions, while dashed lines are unstable solutions. (b-e) Density profiles of $\Psi_{-}$ at different pump amplitudes and frequencies. The color markers shown in (b-e) correspond to dots in (a). Here, $\gamma_{\text{c}}=0.001$ ps$^{-1}$.}
\label{fig:SmallGamma}
\end{figure}

The dynamics of nonequilibrium polariton condensates in the array of microresonator pillars under the resonant pump can be described by the equation~\cite{arrays2}:
\begin{align}\label{model}
i\hbar\frac{\partial\Psi_{\pm}}{\partial t} &=\left[-\frac{\hbar^2}{2m}\triangledown_{\perp}^{2}-i\hbar\frac{\gamma_{\text{c}}}{2}\pm\Omega+g_{\text{c}}|\Psi_{\pm}|^2+V(\textbf{r})\right]\Psi_{\pm} \nonumber \\ 
&+\beta\left(i\partial_{x}\pm\partial_{y}\right)^2\Psi_{\mp}+E_{\pm}(t). 
\end{align}
Here, the indices $\pm$ denote the right-/left-circular polarization components, $m=10^{-4}\times{m_{\text{e}}}$ (${m_{\text{e}}}$ is the free electron mass) is the effective polariton mass, $\gamma_{\text{c}}$ is the polariton loss, $\Omega=0.5$ meV represents the Zeeman splitting caused by an external magnetic field (in this case, the $\Psi_{-}$ component dominates and the $\Psi_{+}$ component is substantially weaker), $g_{\text{c}}=1~\mu \textrm{eV} \cdot \mu \textrm{m}^2$ is the polariton interaction strength that affects the shapes of both polarization components, but not the coupling of them, $\beta=0.25$ meV$\cdot$ $\mu$m$^2$ represents the strength of TE-TM splitting (leading to spin-orbit coupling) intrinsically present in microcavities, $E_{\pm}$ represent the coherent pump, and $V(\textbf{r})=\sum_{\text{m,n}}{\mathcal{V}}(x-x_{\text{m}},y-y_{\text{n}})$ is the potential energy landscape created by the array of microresonator pillars, where ${\mathcal{V}}=V_{0}e^{-[(x-x_{\text{m}})^2+(y-y_{\text{n}})^2]^5/d^{10}}$ describes contribution from individual pillars with the diameter $2d=1.2$ $\mu$m, depth $V_0=-5$ meV, and separation of 1.5 $\mu$m. The pillars are arranged into a finite honeycomb array with a hole in the center, as illustrated in Fig. \ref{fig:EigenStates}(b). The hollow array has two edges, the outer and the inner ones, each with hexagonal shape, so that the width of the ribbon between two edges remains constant. In the simulations, the periodic boundary condition is applied.

To find the edge states supported by this structure, we first analyse the linear, loss-free ($\gamma_{\text{c}}=0$), and pump-free ($E_{\pm}=0$) \eqref{model} by applying the ansatz $\Psi_{\pm}(\textbf{r},t)=u_{\pm}({\textbf{r}})e^{-i{\omega}t}$. The eigenmodes can be obtained by solving the eigenvalue problem: 
\begin{equation}\label{eigenvalue}
{\hbar \omega}u_{\pm}=-\frac{\hbar^2}{2m}\triangledown_{\perp}^{2}u_{\pm}\pm\Omega u_{\pm}+V u_{\pm}+\beta\left(i\partial_{x}\pm\partial_{y}\right)^2u_{\mp}. 
\end{equation}
A part of the dependence of the eigenfrequency $\omega_n$ of the modes on the mode index ($n$) in this structure relevant for edge states is presented in Fig. \ref{fig:EigenStates}(a), where we sorted modes by increasing $\omega_n$. The examples of linear edge states at the outer and inner edges are presented in Figs. \ref{fig:EigenStates}(c) and \ref{fig:EigenStates}(d), respectively. Remarkably, the polaritons excited at the outer edge propagate clockwise, while the polaritons excited at the inner edge propagate counter-clockwise. This is the case for all edge states found within the frequency range corresponding to forbidden topological gap of the infinite array [shaded region in Fig. \ref{fig:EigenStates}(a)]. The states at the outer and inner edges alternate (but in an irregular fashion) in the $\omega_n (n)$ dependence, see example in Fig. \ref{fig:EigenStates}(e). The states outside shaded region correspond to bulk modes.

\begin{figure}[t]
\centering
{\includegraphics[width=1\linewidth]{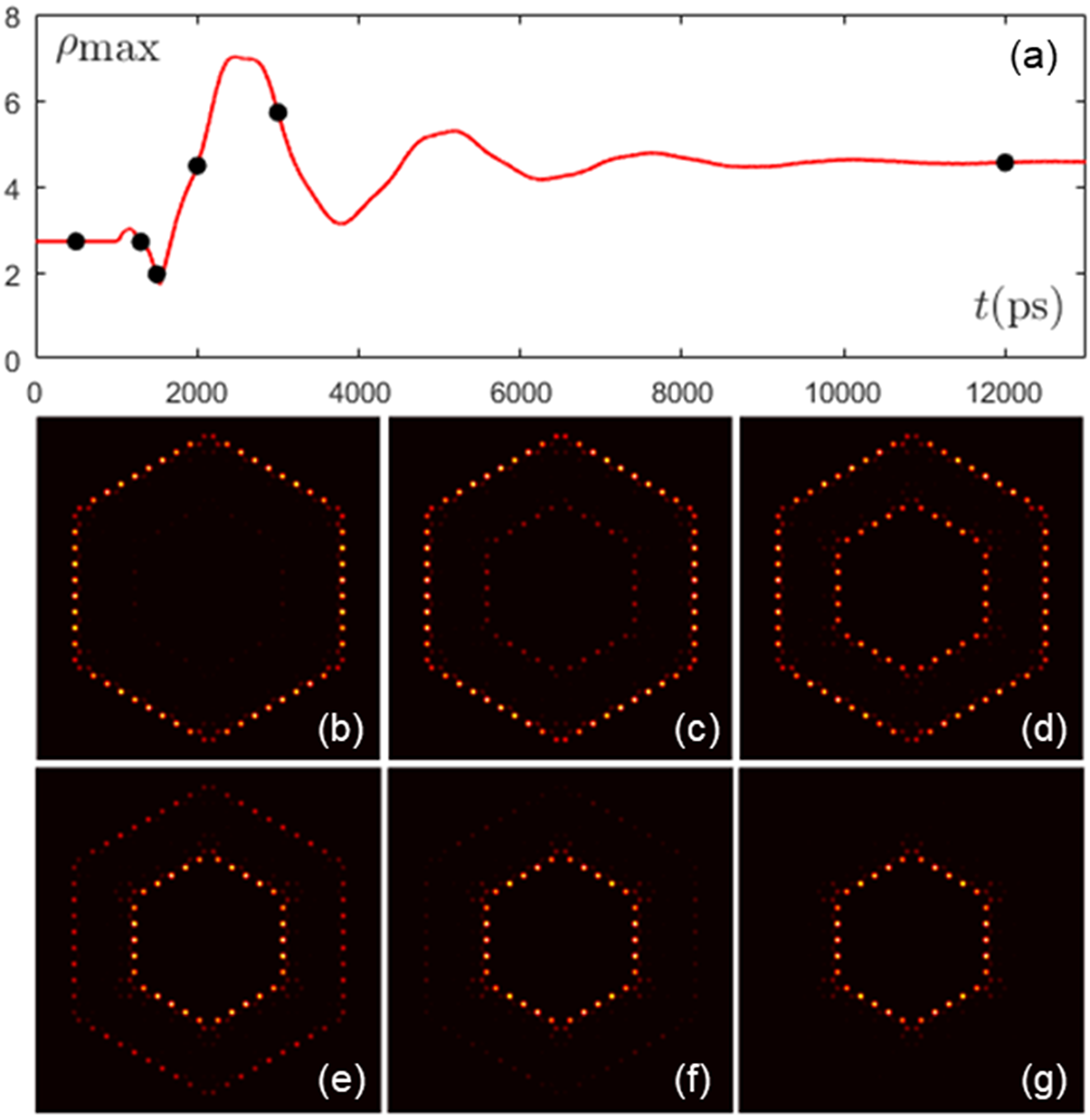}}
\caption{\textbf{Switching between the two edge states.} (a) Time evolution of the peak density ($\rho_\text{max}$ in $\mu$m$^{-2}$) of $\Psi_{-}$ at $E_{\pm}^{0}=0.0001$ with $\omega=-4.3335$ ($t<1000$ ps) and $\omega=-4.3235$ ($t\ge1000$ ps). (b-g) Density profiles of $\Psi_{-}$ at different time moments, corresponding to the black points in (a) from left to right, respectively. Here, $\gamma_{\text{c}}=0.001$ ps$^{-1}$.}
\label{fig:Switching}
\end{figure}

\begin{figure*} [t]
\centering
{\includegraphics[width=1\linewidth]{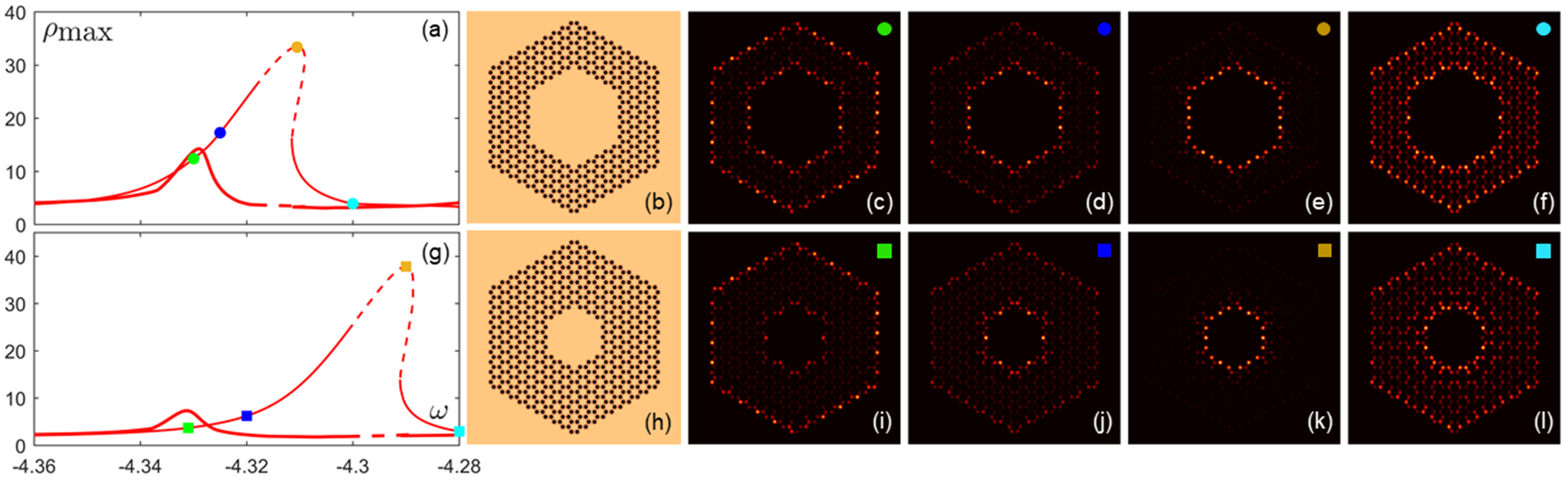}}
\caption{\textbf{Nonlinear edge states at larger loss rate $\gamma_c=0.01$ ps$^{-1}$.} (a) Dependence of the peak density ($\rho_\text{max}$ in $\mu$m$^{-2}$) of $\Psi_{-}$ on the pump frequency ($\omega$ in THz) in (b) honeycomb array with larger hole at $E_{\pm}^{0}=$0.0002 (red lines). (c-f) Density profiles corresponding to the markers in (a). (g) Dependence of the peak density of $\Psi_{-}$ on $\omega$ in (h) honeycomb array with smaller hole at $E_{\pm}^{0}=$0.00015. (i-l) Density profiles corresponding to the markers in (g). In (a,g) the thick (thin) lines show peak density on the outer (inner) edge.}
\label{fig:BigGamma}
\end{figure*}

The excitation of the edge states can be realized using resonant pumps. Here, we consider the plane-wave pump $E_{\pm}=E_{\pm}^{0}e^{-i{\omega}t}$, whose frequency $\omega$ can drive the solution to the desired state. In this case, the stationary edge states in the nonlinear regime can be found by solving the time-independent equation:
\begin{align}\label{stationaryGPE}
\left[-\frac{\hbar^2}{2m}\triangledown_{\perp}^{2}-i\hbar\frac{\gamma_{\text{c}}}{2}\pm\Omega+g_{\text{c}}|u_{\pm}|^2+V(\textbf{r})\right]u_{\pm} \nonumber \\
+\beta\left(i\partial_{x}\pm\partial_{y}\right)^2u_{\mp}+E_{\pm}^{0}-\hbar\omega{u_{\pm}}=0. 
\end{align}

We first assume that the system approaches the conservative regime with a very long polariton lifetime (1000 ps, i.e., $\gamma_{\text{c}}=0.001$ ps$^{-1}$), and the pump is linearly polarized with $E_{+}^{0}=E_{-}^{0}$. Figure \ref{fig:SmallGamma}(a) shows the dependence of the peak density of $\Psi_{-}$ on the frequency and the amplitude of the pump. To check the stability of the edge states, we perturbed the obtained solutions by adding complex (amplitude and phase) broadband noise and let them evolve over long times. The solid (dashed) lines represent the stable (unstable) solutions. Two distinguished resonances can be seen in the frequency domain $\omega=-4.34\sim-4.3$ and each of them corresponds to an edge state. The small $\gamma_\textrm{c}$ is thus required to clearly distinguish two resonances. When the pump amplitude is small [see the blue line in Fig. \ref{fig:SmallGamma}(a)], the two peaks are almost independent with the left resonance corresponding to strongly excited outer edge and practically unexcited inner edge [Fig. \ref{fig:SmallGamma}(b)], while in the right resonance the situation is inverted, and polaritons concentrate exclusively on the inner edge as shown in Fig. \ref{fig:SmallGamma}(c). Increasing the pump amplitude strengthens the coupling of the two edge states [see the red lines in Fig. \ref{fig:SmallGamma}(a)], since corresponding resonances broaden and start to partially overlap. As a result, in Fig. \ref{fig:SmallGamma}(d) one can observe a mixed state containing comparable contributions from both inner and outer edges of the structure. With further increase in frequency [Fig. \ref{fig:SmallGamma}(e)] the nonlinear edge state localizes at the inner edge. Thus, even at such large lifetimes, nonlinearity substantially affects the structure and location of the edge states in the hollow honeycomb structure.

\begin{figure}[t]
\centering
{\includegraphics[width=1\linewidth]{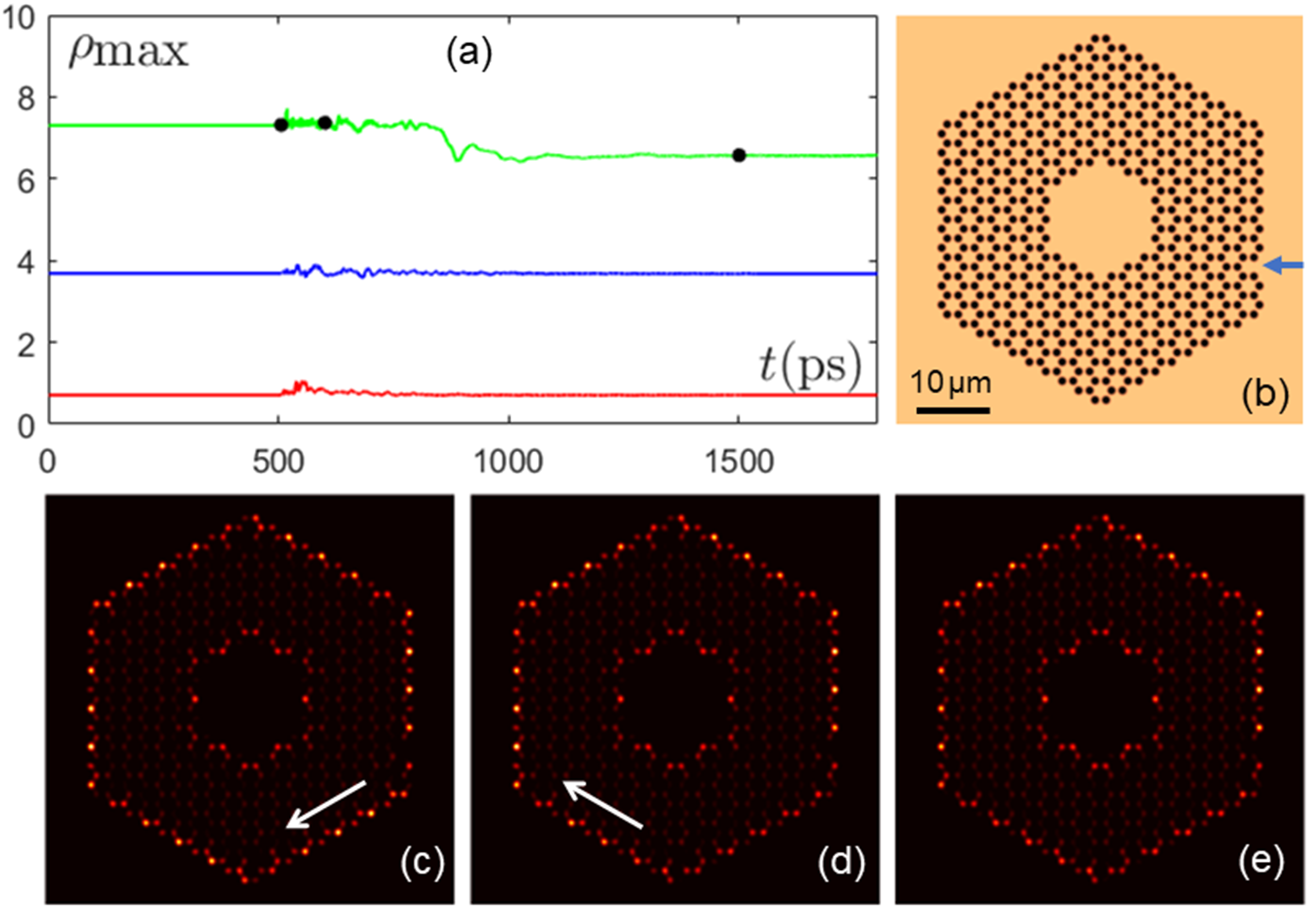}}
\caption{\textbf{Defect-induced coupling of two edge states.} (a) Time evolution of the peak density ($\rho_\text{max}$ in $\mu$m$^{-2}$) of $\Psi_{-}$ at the outer edge (green line), inner edge (blue line), and in the bulk (red line). The potential at $t<500$ ps is shown in Fig. 4(h) and the potential with a missing pillar, indicated by the blue arrow, is shown in (b). (c-e) Density profiles at different time scales in (a), corresponding to the black points from left to right, respectively. The white arrows indicate the polariton current at the outer edge. The density profile at $t<500$ ps is shown in Fig. 4(i). Here, $\gamma_c=0.01$ ps$^{-1}$.}
\label{fig:MissingPillar}
\end{figure}

Due to the nonequilibrium nature of polaritons, one can realize switching between two edge states by suddenly varying the frequency of the pump, as shown in Fig. \ref{fig:Switching}. When $t<1000$ ps, the pump frequency is $\omega=-4.3335$ and the outer edge is excited [Fig. \ref{fig:Switching}(b)]. If we suddenly increase the pump frequency at $t=1000$ ps to $\omega=-4.3235$, that is close to the eigenfrequency of the inner edge state, as shown in Figs. \ref{fig:SmallGamma}(a) and \ref{fig:SmallGamma}(c), the polaritons start to occupy the inner edge [Fig. \ref{fig:Switching}(c)] quickly reaching the same density as at the outer edge [Fig. \ref{fig:Switching}(d)]. After that, the polaritons at the outer edge decay very slowly [Figs. \ref{fig:Switching}(a,e,f)] and finally disappear, leaving only the inner edge occupied [Figs. \ref{fig:Switching}(a,g)]. Even though the whole switching process takes a very long time $\sim$10 ns, it is just 10 times longer than the polariton lifetime. This ratio takes place also for much smaller polariton lifetimes, where transitions to a new steady state after changing the parameters of the structure occurs much faster, an example of which is presented in Fig. \ref{fig:MissingPillar}. This mechanism of switching is very robust in comparison with previously suggested schemes based on array modulations \cite{resonant1}.

For larger polariton loss rates the resonance peaks in the dependence on frequency $\omega$ substantially broaden as shown in Fig. \ref{fig:BigGamma}(a), where we show polariton densities at the different edges separately to distinguish the two peaks. The thick (thin) lines represent the maximum density at the outer (inner) edge. In this case, the coupling of the two edge states becomes much stronger [Figs. \ref{fig:BigGamma}(c,d)], and solutions become more stable. Increased coupling between the two edge states leads to their coexistence, especially when the outer edge is predominantly occupied [Fig. \ref{fig:BigGamma}(c)]. To decouple the two edge states and to increase the separation between corresponding resonances to make them more distinguishable, one can reduce the size of the central hole, increasing the width of the ribbon between inner and outer edges. This is done in Fig. \ref{fig:BigGamma}(h) where the outer edge is unchanged but the inner one is substantially reduced. As a consequence, the peak density of the state on the inner edge increases, nonlinearity results in more pronounced shift of the tip of corresponding resonance curve to the right, so that the separation between two resonances increases, as shown in Fig. \ref{fig:BigGamma}(g). In Figs. \ref{fig:BigGamma}(i,j) the contrast between states residing at two different edges becomes higher. For both structures, due to nonlinear tilt of right resonance in Figs. \ref{fig:BigGamma}(a,g), in the tip of this resonance the state on the inner edge clearly dominates [Figs. \ref{fig:BigGamma}(e,k)]. At the right side of the right peaks in Figs. \ref{fig:BigGamma}(a,g) (see the cyan markers) the solutions in Figs. \ref{fig:BigGamma}(f,l) show that more polaritons lie between the two edges, although the edges are mainly occupied. Corresponding frequency values are close to the border of the topological gap, hence corresponding states deeply penetrate into the structure.

The influence of the polarization of the pump on the edge states was studied too. Switching the pump from linearly polarized ($E_{+}^{0}=E_{-}^{0}$) to circularly polarized ($E_{+}^{0}=$0 and $E_{-}^{0}=$0.0002) results only in a slight decrease of peak amplitudes of corresponding states. However, for pump with opposite circular polarization $E_{+}^{0}=$0.0002 and $E_{-}^{0}=$0 the excitation efficiency is very low and the system operates in quasi-linear regime.

To demonstrate that the edge states are topologically protected, i.e. they cannot be destroyed or scattered to the bulk by a potential defect, we erase an arbitrary potential pillar as indicated by the blue arrow in Fig. \ref{fig:MissingPillar}(b). The target pillar can also be addressed in an all-optical manner, using an off-resonant beam focused on the target pillar, that can load condensed polaritons into it affecting nearby polariton currents~\cite{MaNatComm}. The steady currents forming at $t<500$ ps, before the pillar is removed, are shown in Fig. \ref{fig:MissingPillar}(c). The disappearance of the selected pillar at  $t>500$ ps does not lead to scattering into bulk [see the red line in Fig. \ref{fig:MissingPillar}(a) showing amplitude in the bulk] or counter-clockwise polariton propagation [see the white arrows in Figs. \ref{fig:MissingPillar}(c-e)]. Removal of the pillar practically does not affect the amplitude at the inner edge, even though the amplitude on the outer edge slightly reduces after some time [Fig. \ref{fig:MissingPillar}(a)]. 

In summary, we have demonstrated that hollow honeycomb arrays support topological polariton currents at the outer and inner edges that are coupled by nonlinearity and that can be switched on and off in a controllable fashion by varying the frequency of the pump in this dissipative system.

\textbf{Funding.} Deutsche Forschungsgemeinschaft (DFG) (No. 231447078, 270619725); Paderborn Center for Parallel Computing, PC$^2$; Russian Science Foundation (Project 17-12-01413-$\Pi$); Spanish MINECO through Project No. TEC2017-86102-C2-1-R.

\textbf{Disclosures.} The authors declare no conflicts of interest.


\begin{thebibliography}{99}

\bibitem{elect1}
M. Z. Hasan and C. L. Kane,
Rev. Mod. Phys. \textbf{82}, 3045 (2010).

\bibitem{elect2}
X.-L. Qi and S.-C. Zhang,
Rev. Mod. Phys. \textbf{83}, 1057 (2011).

\bibitem{photon1}
L. Lu, J. D. Joannopoulos, and M. Soljacic,
Nat. Photon. \textbf{8}, 821 (2014).

\bibitem{photon2}
T. Ozawa, H. M. Price, A. Amo, N. Goldman, M. Hafezi, L. Lu, M. C. Rechtsman, D. Schuster, J. Simon, O. Zilberberg, and I. Carusotto,
Rev. Mod. Phys. \textbf{91}, 015006 (2019).

\bibitem{giro1}
F. D. M. Haldane and S. Raghu,
Phys. Rev. Lett. \textbf{100}, 013904 (2008).

\bibitem{giro2}
Z. Wang, Y. Chong, J. D. Joannopoulos, and M. Soljacic,
Nature \textbf{461}, 772 (2009).

\bibitem{reson1}
M. Hafezi, E. A. Demler, M. D. Lukin, and J. M. Taylor,
Nat. Phys. \textbf{7}, 907 (2011).

\bibitem{reson2}
O. Umucalilar and I. Carusotto,
Phys. Rev. Lett. \textbf{108}, 206809 (2012).

\bibitem{floq1}
N. H. Lindner, G. Refael, and V. Galitski,
Nat. Phys. \textbf{7}, 490 (2011).

\bibitem{floq2}
M. C. Rechtsman, J. M. Zeuner, Y. Plotnik, Y. Lumer, D. Podolsky, F. Dreisow, S. Nolte, M. Segev, and A. Szameit,
Nature \textbf{496}, 196 (2013).

\bibitem{toppol1}
A. V. Nalitov, D. D. Solnyshkov, and G. Malpuech,
Phys. Rev. Lett. \textbf{114}, 116401 (2015).

\bibitem{toppol2}
T. Karzig, C.-E. Bardyn, N. H. Lindner, and G. Refael,
Phys. Rev. X \textbf{5}, 031001 (2015).

\bibitem{toppol3}	
C.-E. Bardyn, T. Karzig, G. Refael, and T. C. H. Liew,
Phys. Rev. B \textbf{91}, 161413 (2015).

\bibitem{toppol4}
S. Klembt, T. H. Harder, O. A. Egorov, K. Winkler, R. Ge, M. A. Bandres, M. Emmerling, L. Worschech, T. C. H. Liew, M. Segev, C. Schneider, and S. Höfling,
Nature \textbf{562}, 552 (2018).

\bibitem{toppol5}
P. St-Jean, V. Goblot, E. Galopin, A. Lemaitre, T. Ozawa, L. Le Gratiet, I. Sagnes, J. Bloch, A. Amo,
Nat. Photon. \textbf{11}, 651 (2017).

\bibitem{nonltop1}
D. Smirnova, D. Leykam, Y. D. Chong, and Y. Kivshar,
Appl. Phys. Rev. \textbf{7}, 021306 (2020).

\bibitem{topsol1}
Y. Lumer, Y. Plotnik, M. C. Rechtsman, and M. Segev,
Phys. Rev. Lett. 111, 243905 (2013).

\bibitem{topsol2}
M. J. Ablowitz, C. W. Curtis, and Y.-P. Ma,
Phys. Rev. A \textbf{90}, 023813 (2014).

\bibitem{topsol3}
D. Leykam and Y. D. Chong,
Phys. Rev. Lett. \textbf{117}, 143901 (2016).

\bibitem{topsol4}
S. K. Ivanov, Y. V. Kartashov, A. Szameit, L. Torner, and V. V. Konotop,
ACS Photonics \textbf{7}, 735 (2020).

\bibitem{topsol5}
S. Mukherjee and M. C. Rechtsman,
Science \textbf{368}, 856 (2020).

\bibitem{topsol6}
Y. V. Kartashov and D. V. Skryabin,
Optica \textbf{3}, 1228 (2016).

\bibitem{topsol7}
D. R. Gulevich, D. Yudin, D. V. Skryabin, I. V. Iorsh, and I. A. Shelykh,
Sci. Rep. \textbf{7}, 1780 (2017).

\bibitem{topsol8}
C. Li, F. Ye, X. Chen, Y. V. Kartashov, A. Ferrando, L. Torner, and D. V. Skryabin,
Phys. Rev. B \textbf{97}, 081103 (2018).

\bibitem{nonlpol1}
O. Bleu, D. D. Solnyshkov, and G. Malpuech,
Phys. Rev. B \textbf{93}, 085438 (2016).

\bibitem{nonlpol2}
C.-E. Bardyn, T. Karzig, G. Refael, and T. C. H. Liew,
Phys. Rev. B \textbf{93}, 020502 (2016).

\bibitem{nonlpol3}
R. Banerjee, S. Mandal, and T. C. H. Liew,
Phys. Rev. Lett. \textbf{124}, 063901 (2020).

\bibitem{bistab1}
Y. V. Kartashov and D. V. Skryabin,
Phys. Rev. Lett. \textbf{119}, 253904 (2017).

\bibitem{bistab2}
W. F. Zhang, X. F. Chen, Y. V. Kartashov, D. V. Skryabin, and F. W. Ye,
Laser Photon. Rev. \textbf{13}, 1900198 (2019).

\bibitem{arrays1}
V. G. Sala, D. D. Solnyshkov, I. Carusotto, T. Jacqmin, A. Lema\^{\i}tre, H. Ter\c{c}as, A. Nalitov, M. Abbarchi, E. Galopin, I. Sagnes, J. Bloch, G. Malpuech, and A. Amo, 
Phys. Rev. X \textbf{5}, 011034 (2015).

\bibitem{arrays2}
C. Schneider, K. Winkler, M. D. K.  Fraser, M. Kamp, Y. Yamamoto, E. A.  Ostrovskaya, and S. Höfling,  
Rep. Prog. Phys. \textbf{80}, 016503 (2017).

\bibitem{honey1}
M. Milicevic, T. Ozawa, P. Andreakou, I. Carusotto, T. Jacqmin, E. Galopin, A. Lemaitre, L. Le Gratiet, I. Sagnes, J. Bloch, and A. Amo,
2D Mat. \textbf{2}, 034012 (2015).

\bibitem{lieb1}
C. E. Whittaker, E. Cancellieri, P. M. Walker, D. R. Gulevich, H. Schomerus, D. Vaitiekus, B. Royall, D. M. Whittaker, E. Clarke, I. V. Iorsh, I. A. Shelykh, M. S. Skolnick, and D. N. Krizhanovskii,
Phys. Rev. Lett. \textbf{120}, 097401 (2018).

\bibitem{kagome1}
D. R. Gulevich, D. Yudin, I. V. Iorsh, and I. A. Shelykh,
Phys. Rev. B \textbf{94}, 115437 (2016).

\bibitem{resonant1}
Y. Q. Zhang, Y. V. Kartashov, Y. P. Zhang, L. Torner, and D. V. Skryabin,
Laser Photon. Rev. \textbf{12}, 1700348 (2018).

\bibitem{MaNatComm}
X. Ma, B. Berger, M. A{\ss}mann, R. Driben, T. Meier, C. Schneider, S. H\"{o}fling, and S. Schumacher,
Nat. Commun. \textbf{11}, 897 (2020).

\end{thebibliography}
\end{document}